# Centrifugal deterministic lateral displacement separation system

Mingliang Jiang, Aaron D. Mazzeo and German Drazer

*Department of Mechanical and Aerospace Engineering,
Rutgers, The State University of New Jersey, Piscataway, NJ 08854*

**Abstract**

This work investigates the migration of spherical particles of different sizes in a centrifuge-driven deterministic lateral displacement (c-DLD) device. Specifically, we use a scaled-up model to study the motion of suspended particles through a square array of cylindrical posts under the action of centrifugation. Experiments show that separation of particles by size is possible depending on the orientation of the driving acceleration with respect to the array of posts (forcing angle). We focus on the fractionation of binary suspensions and measure the separation resolution at the outlet of the device for different forcing angles. We found excellent resolution at intermediate forcing angles, when large particles are locked to move at small migration angles but smaller particles follow the forcing angle more closely. Finally, we show that reducing the initial concentration (number) of particles, approaching the dilute limit of single particles, leads to increased resolution in the separation.

**Introduction**

Centrifuge-based microfluidic systems are an emerging lab-on-a-chip platform that has attracted significant interest within the analytical sciences in general (Vázquez et al. 2011), and for biological and biomedical applications in particular, including *in vitro* diagnostics (Gorkin et al. 2010), high-throughput screening in drug discovery (Madou et al. 2006), as well as various cell manipulation, sorting and analysis applications (Burger et al. 2012). These *lab-on-a-CD* or *lab-on-a-disc* systems (Madou et al. 2006), offer unique features and potential advantages compared to alternative lab-on-a-chip methods of driving flow and particles (Madou et al. 2006; Ducrée et al. 2007). First, both fluid and/or particle motion can be achieved based on the rotation of the system and require, in principle, no additional actuation elements, such as external syringe pumps (a detailed comparison of actuation methods is provided in the review by Madou et al. (2006)). This independence from external pumps facilitates the design of fully enclosed, low-cost, and disposable chips (Gorkin et al. 2010; Burger et al. 2012). The use of closed systems is advantageous in potentially reducing risks of contamination (Madou et al. 2006). Additionally, it has also been recognized that, compared to other on-chip driving methods, such as electroosmotic pumping, centrifugal actuation is less dependent on fluid physicochemical properties, which provides much needed adaptability in the handling of biological samples that could vary significantly from sample to sample (Ducrée et al. 2007). More recently, in an effort to simplify access to these technologies, the concepts of *microfluidic apps* (Mark et al. 2012) and *LabTube* cartridges (Kloke et al. 2014) were introduced. In both cases, the underlying concept is to develop inexpensive, disposable micro-total analysis systems (μ-TAS) that can be used with standard laboratory centrifuges. In all cases, a unit operation at the core of the majority of applications, especially those dealing with the manipulation of cells or other particles, is a separation or sorting unit.

Here, we show that deterministic lateral displacement (DLD), a popular and promising microfluidic separation method, which was originally based on a pressure driven flow field, can be successfully implemented as a centrifuge-driven technique. The proposed centrifugal DLD (c-DLD) device retains advantages of traditional, flow-driven DLD devices (McGrath et al. 2014), including its intrinsic two-dimensional and continuous separation capabilities (*vector chromatography*), and can be easily integrated into lab-on-a-CD systems. On the other hand, c-DLD has a significant advantage over the flow driven case: the possibility to control the relative orientation of the driving field with respect to the ordered array of cylindrical obstacles, which is the critical parameter that determines the resulting separation process. By contrast, in order to change the flow direction with respect to the obstacle array in traditional DLD would require the design and fabrication of an entirely new device. We take advantage of the ability to (easily) control the direction of the obstacle array with respect to centrifugal acceleration to investigate a range of forcing angles and to identify conditions leading to separation for suspended particles of different size. In all cases, we use a millimeter-sized system (scaled-up version of a microdevice), track the individual trajectory of the particles as they move through the array, and analyze the distribution of particles at the outlet and the resolution of the separation at different angles. We also investigate the limitations of the system by considering the release of increasingly large groups of particles with its associated hindrance on the resolution.

## Deterministic lateral displacement separation systems

Deterministic lateral displacement is a separation method based on the different migration angles exhibited by suspended particles of different size as they move through an ordered array of cylindrical posts (*obstacles*) (McGrath et al. 2014). It was originally introduced as a *passive* method in which a flow field drives a mixture of particles through a square array of obstacles. The obstacle array is slanted with respect to the average flow field and, as a result, particles move in different directions depending on their size (Huang et al. 2004). More recently, it has been shown that DLD can also fractionate samples based on shape and deformability (Quek et al. 2011; Bogunovic et al. 2012; Beech et al. 2012; Zeming et al. 2013; Ranjan et al. 2014; Krüger et al. 2014; Holmes et al. 2014). DLD has been successful in the fractionation of various mixtures, including biological samples and cells in particular (Beech et al. 2012; Zeming et al. 2013; Ranjan et al. 2014; Krüger et al. 2014; Holmes et al. 2014; Siyang Zheng et al. 2005; Davis et al. 2006; Nan Li et al. 2007; Huang et al. 2008; Inglis et al. 2008; Morton et al. 2008; Green et al. 2009; Inglis et al. 2010; Holm et al. 2011; Al-Fandi et al. 2011; Inglis et al. 2011; Zhang et al. 2012; Loutherback et al. 2012; Liu et al. 2013; Ozkumur et al. 2013; Karabacak et al. 2014), which makes it attractive for lab-on-a-CD applications. In addition, different post geometries (Loutherback et al. 2010; Al-Fandi et al. 2011; Ranjan et al. 2014) and other design considerations (Inglis 2009) have been proposed to improve performance.

In previous work, we showed that DLD could also be operated as an *active* method using external forces to drive the separation. In particular, we demonstrated the use of both gravity and electrokinetic flow to separate suspended particles in microfluidic DLD systems (Devendra and Drazer 2012; Hanasoge et al. 2014). We also showed that both driving fields lead to transport and fractionation analogous to the flow-driven case, and in excellent agreement with our previous studies performed in scaled-up versions of microfluidic systems (Balvin et al. 2009; Bowman et al. 2012; Bowman et al. 2013). Here, we also use a scaled-up microfluidic system that allows us to explore a broad range of working conditions, identify those that lead to separation, and capture resolution trends depending on initial conditions.

A schematic view of the array of obstacles used in DLD systems is presented in Figure 1. We define the forcing angle $\theta$, as the angle between the centrifugal acceleration (indicated by the solid line $C$ in the figure) and a principal direction in the square array (indicated by the dashed line $A$ in the figure). We also define the migration angle $\alpha$ of a given type of particles as the average angle of their trajectory with respect to the principal direction of the array (see solid line $M$ in Figure 1). In previous modeling work, we have shown that for each species, there is a critical value of the forcing angle, $\theta_c$, such that, for forcing angles smaller than the critical value, particles will be locked to move along a column in the device with $\alpha=0°$ (Frechette and Drazer 2009; Herrmann et al. 2009; Risbud and Drazer 2013; Risbud and Drazer 2014). We also show in Figure 1a) that, in general, the critical angle increases with particle size and, as a result, larger particles remain locked to move along the principal direction of the array for larger forcing angles. Note that the angle between the centrifugal acceleration and the array of obstacles will change continuously depending on the position within the array (see a schematic view in Figure 1b). The magnitude of the variation in the forcing angle within the array depends on the size of the array with respect to the size of the centrifuge. Here, for simplicity, we define the forcing angle as the angle between the array and a radial force that goes through the center of the array (as shown in Figure 1a). The objective is to investigate conditions that will lead to the situation schematically shown in Figure 1a, which corresponds to a clear separation between small and large particles. In fact, this work focuses on the fractionation of binary suspensions.

## Experimental setup

We perform experiments using *meso models* of DLD microfluidic systems, that is, we scale-up a DLD microdevice from micrometer to millimeter scale. For example, instead of individual posts with diameters of 10 μm, we use posts that are 1 mm in diameter. Of course, for the meso models to reproduce the behavior at the micro scale we need to maintain the dimensionless parameters that characterize the transport process constant. First, the geometric similarity is guaranteed by preserving the aspect ratios between the array of obstacles and the size of the different particles to be separated. Second, dynamic similarity is maintained by using a higher viscosity fluid, as detailed below. Finally, we take advantage of the fact that the separation mechanism is deterministic, and thus independent of Brownian motion. Therefore, the only condition on the Peclet number is that it is large enough so that Brownian motion can be ignored, which is clearly satisfied with millimeter-scale particles. Using such macroscopic systems simplifies several aspects of the experimental work significantly, including fabricating the device, in which all parts can be 3D printed, adjusting the external forcing angle with a simple rotation of the device, re-using the system in multiple experiments at different angles and with different mixtures of particles and capturing the trajectory followed by the particles using a high-speed digital camera (in combination with a tachometer system to synchronize the acquisition of snapshots).

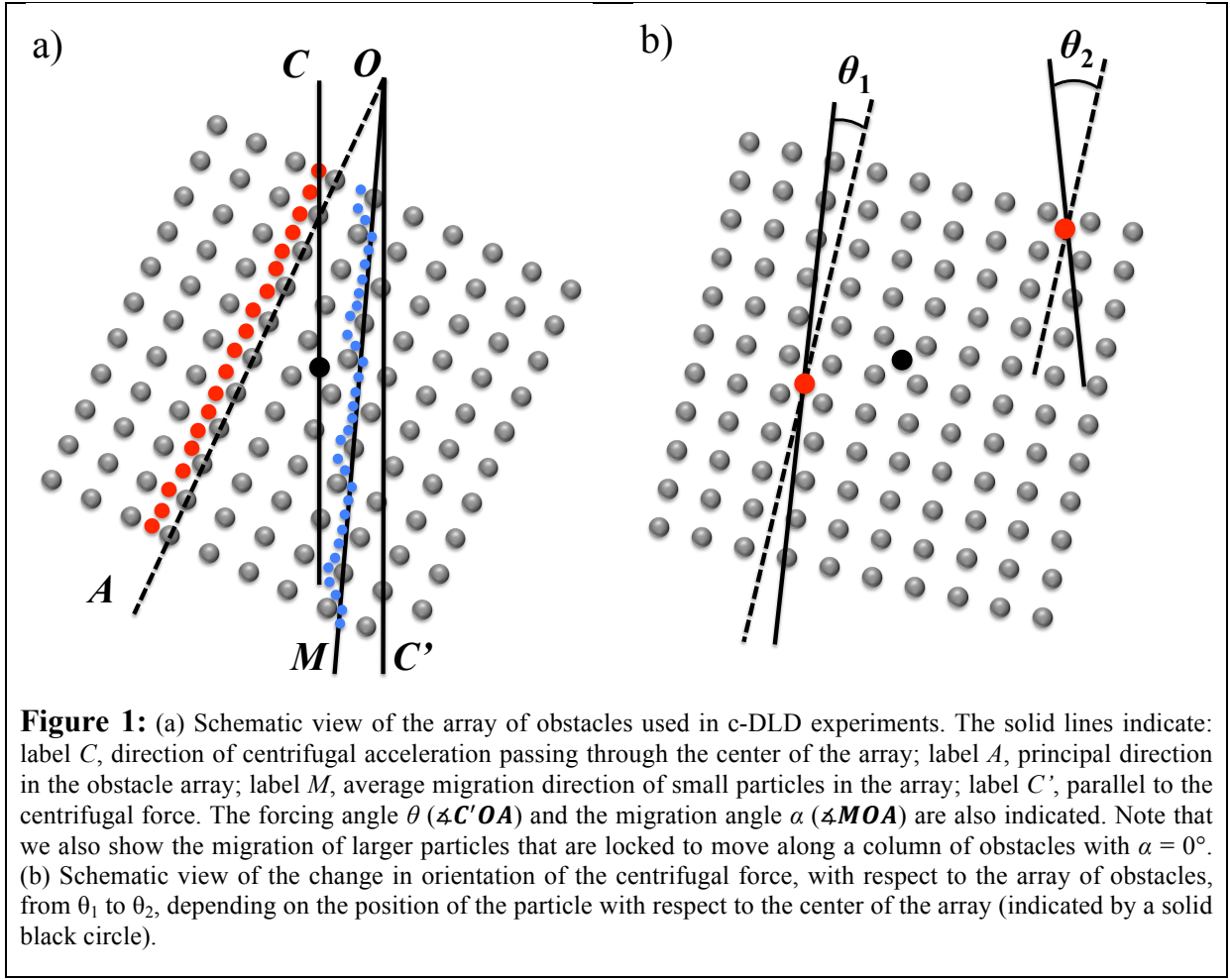

**Figure 1:** (a) Schematic view of the array of obstacles used in c-DLD experiments. The solid lines indicate: label *C*, direction of centrifugal acceleration passing through the center of the array; label *A*, principal direction in the obstacle array; label *M*, average migration direction of small particles in the array; label *C'*, parallel to the centrifugal force. The forcing angle $\theta$ ($\angle C'OA$) and the migration angle $\alpha$ ($\angle MOA$) are also indicated. Note that we also show the migration of larger particles that are locked to move along a column of obstacles with $\alpha = 0°$. (b) Schematic view of the change in orientation of the centrifugal force, with respect to the array of obstacles, from $\theta_1$ to $\theta_2$, depending on the position of the particle with respect to the center of the array (indicated by a solid black circle).

The centrifugal DLD experimental setup is shown schematically in Figure 2 and is composed of four parts. A base and rotating board are first attached to the rotating plate of the centrifuge and fixed in place. A housing for the separation device is then mounted on the rotating board. This housing can be rotated with respect to the base and thus controls the forcing angle, as shown in Figure 2a, in intervals of 5 degrees over the entire range of possible orientations. Then, the actual separation system can be installed inside the housing and filled with fluid. Particles are then introduced in the inlet region and a transparent cover is used to close the device. All parts are fabricated using a standard 3D printer. The centrifuge has a steady-state rotation of 600 rpm (Jouan RC1010).

We use two types of particles: cellulose acetate particles (1.26 g/cm$^3$) with diameters of 1.00 mm and 1.50 mm and Delrin® acetal particles (1.41 g/cm$^3$) with diameters of 1.59 mm and 2.38 mm. In previous work, we have shown that posts smaller than the particles typically lead to a higher resolution (Bowman et al. 2013). In addition, in preliminary experiments, we observed that particles smaller than the obstacles would sometimes get trapped in front of posts and would not come out of the array. Based on these considerations, we decided to use relatively thin posts with a diameter of 0.5 mm (and 3 mm in height). The size of the entire device is 59 mm by 39 mm. The DLD lattice contains 12 by 10 cylindrical obstacles. The cylindrical posts are separated by 3.5 mm center-to-center. In order to reduce the operating Reynolds numbers, we use a 50% by volume mixture of water and glycerol, with an estimated viscosity $\mu = 6.9$ mPa·s and density $\rho = 1.14$ g/cm$^3$. The Reynolds numbers based on particle velocity are O(1), which is particularly relevant for separation microdevices working at high throughput. Compared to standard microfluidic DLD systems, the current device is approximately scaled up by a factor between 10 and 100.

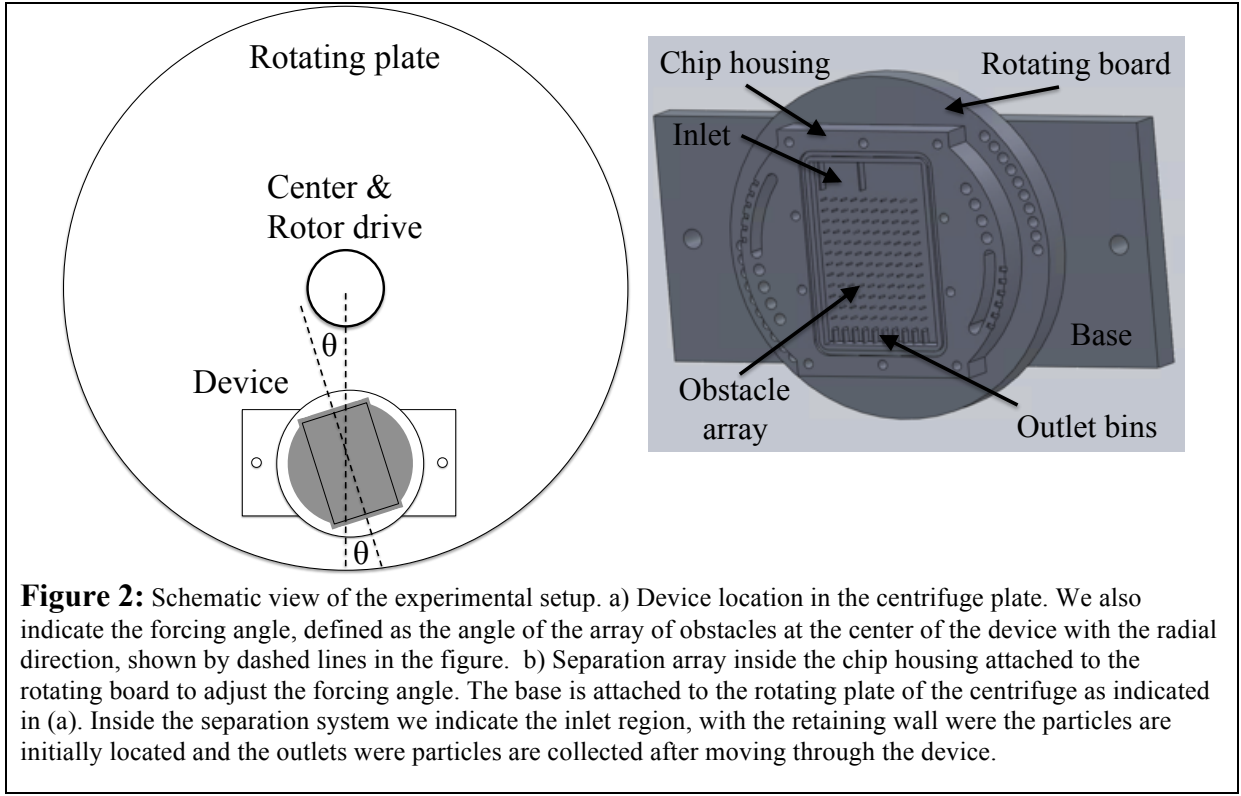

**Figure 2:** Schematic view of the experimental setup. a) Device location in the centrifuge plate. We also indicate the forcing angle, defined as the angle of the array of obstacles at the center of the device with the radial direction, shown by dashed lines in the figure. b) Separation array inside the chip housing attached to the rotating board to adjust the forcing angle. The base is attached to the rotating plate of the centrifuge as indicated in (a). Inside the separation system we indicate the inlet region, with the retaining wall were the particles are initially located and the outlets were particles are collected after moving through the device.

In all the experiments, we release an equal number of particles of two different sizes from the inlet region. A small retaining wall guides the particles toward the array of obstacles at the beginning of the experiments. We do not use valves or any other control system to release the particles, and they move freely throughout the experiments, including the transient before steady state rotation is reached. After moving through the array particles are collected in 10 bins fabricated at the end of the device. The results are in fact reported in terms of bin number from 0 (corresponding to 0° migration angle) to 10 (corresponding to 35° the largest possible migration angle in the array). We only consider particles that reach the outlet and discard a small number of events in which a particle is trapped inside the array (less than 10% of the particles). We also investigate the effect that the initial number of particles released together has on the separation resolution. In general, we use the following measure of resolution,

$$R = \frac{\mu_s - \mu_b}{2(\sigma_s + \sigma_b)}, \qquad (1)$$

where $\mu_s$, $\mu_b$ are the average outlet number for small and large particles, respectively, and $\sigma_s$, $\sigma_b$ are the corresponding standard deviation in the outlet number.

## Results and discussion

First, we performed experiments for increasing forcing angles for a mixture of Delrin® acetal particles (1.59 mm and 2.38 mm in diameter). Specifically, we measured the distribution of particles over the 10 outlets for forcing angles 15°, 20° and 25° by performing repeated experiments in which we initially placed two small and two large particles in the inlet of the device (see Figure 3). It is clear from Figure 3 that the distribution of larger particles at the outlet does not change significantly as the forcing angle is increased and most of the particles are collected at outlets 1 and 2. This is consistent with a situation in which the critical angle for the 2.38 mm particles is larger than the highest forcing angle considered here and, as a result, the particles essentially remain locked to move along the first couple of columns of obstacles in the device, as schematically shown in Figure 1a. On the other hand, even at the smallest forcing angle, some of the small particles migrate towards higher outlet numbers. Moreover, at the largest forcing angle particles migrate to the opposite corner of the device, with an average outlet number slightly higher than 9. This is also consistent with our previous results, indicating that smaller particles have smaller critical angles. In summary, the resolution increases continuously from R=0.21 for a forcing angle of 15° to an excellent resolution R=1.39 for $\theta = 25°$. A significant difference with our previous work is the fact that we observe a relatively broad distribution of particles, probably due to dispersion in the initial position of the particles in the inlet area, which changes from experiment to experiment, contributing to the observed *noise* in the outlet distribution. Another important difference is that particles are introduced at the same time, and not individually as in our previous work. This could lead to particle-particle interactions, which could affect the trajectory of the individual particles, as we discuss below. We note however that the current method of introducing the particles is likely to provide more realistic results in terms of actual separation experiments.

In Figure 4, we present snapshots at time intervals of approximately 1 second, corresponding to an experiment in which we separated Delrin® acetal particles at a forcing angle $\theta = 25°$. Clearly, large particles are not able to move across columns of obstacles and are locked at nearly zero migration angles. On the other hand, smaller particles migrate to the opposite corner of the device, resulting in excellent separation resolution.

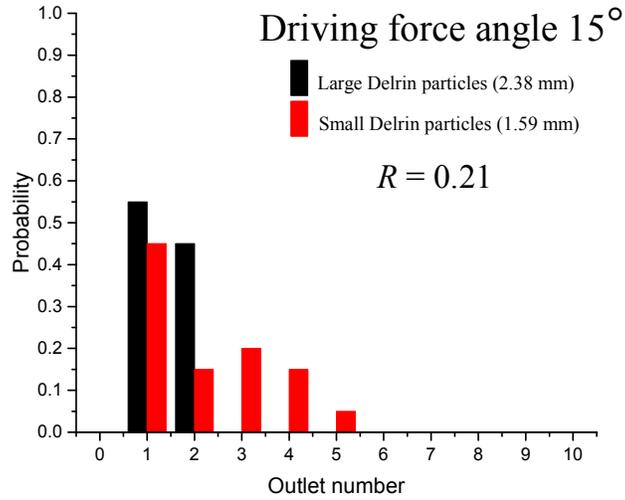
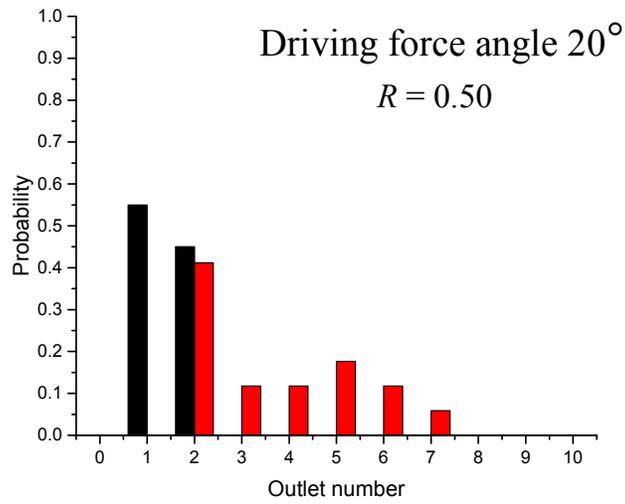
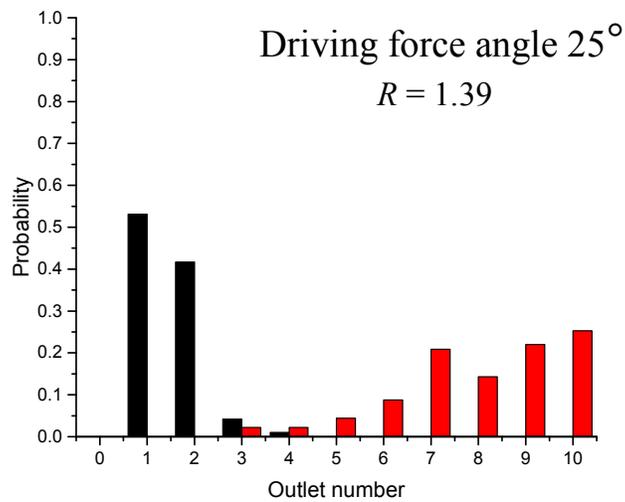

**Figure 3** Probability distributions of particles at the outlet for different forcing angles, as indicated. The sample is a binary mixture of Delrin® acetal particles (1.59 mm and 2.38 mm in diameter).

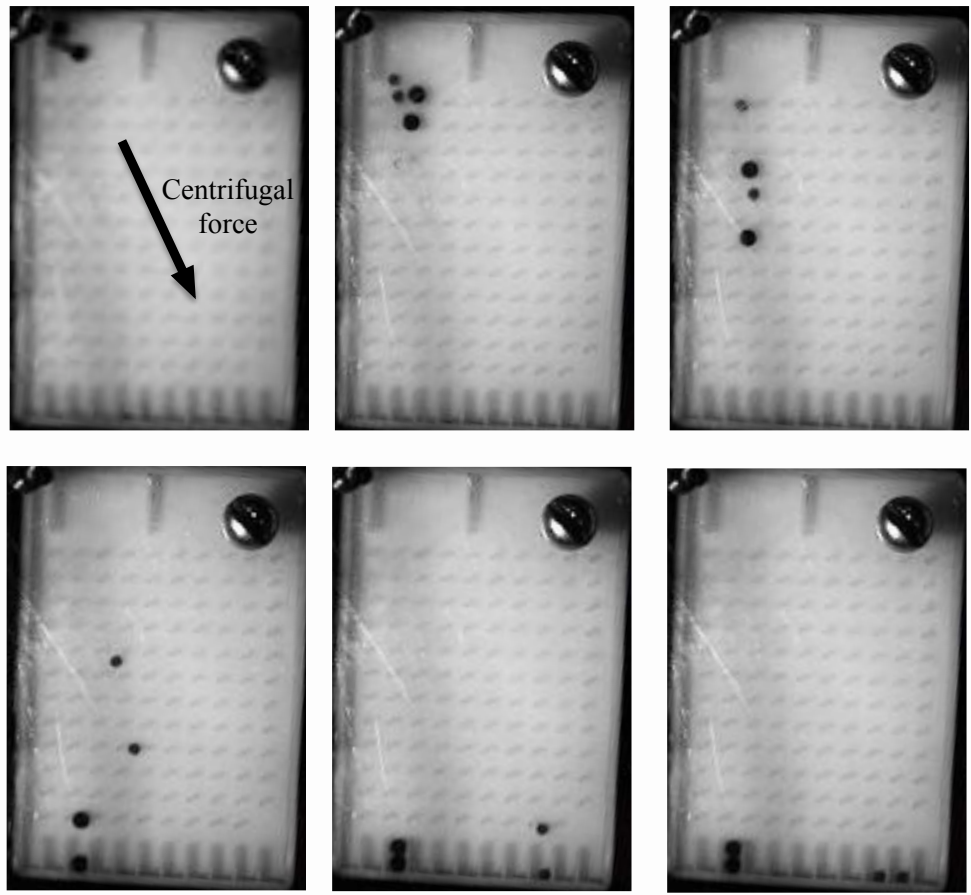

**Figure 4:** Snapshots at approximately 1 second time intervals during the separation of Delrin® acetal particles at a forcing angle $\theta = 25°$. We indicate the forcing angle with an arrow in the first snapshot.

Then, we studied the transport of a binary mixture of smaller particles but with the same relative size difference (cellulose acetate particles of 1.0 mm and 1.5 mm in diameter). Specifically, we considered the same forcing angles as before, $\theta = 15°$, $20°$ and $25°$, and measured the distribution of particles at the outlet as shown in Figure 5. First of all, we note that in all cases the resolution is significantly reduced. At the smallest forcing angle, all the particles are significantly confined to small migration angles and they are mostly distributed over outlets 1-3, resulting in low resolution. Note that the migration angle or average outlet is smaller than those obtained with the larger acetal particles, suggesting a smaller critical angle, which is not expected following our previous work. However, the differences are small compared to the variance in the distributions. The resolution improves at intermediate driving angles, $R = 0.46$, mainly because the smaller particles migrate at larger angles. However, upon increasing the angle further, all particles migrate significantly thus decreasing the resolution again. In addition, the distribution of particles broadens and the outlet numbers range from 4 to 10. The resulting resolution is, by chance, exactly zero, due to equal average outlet for both particles.

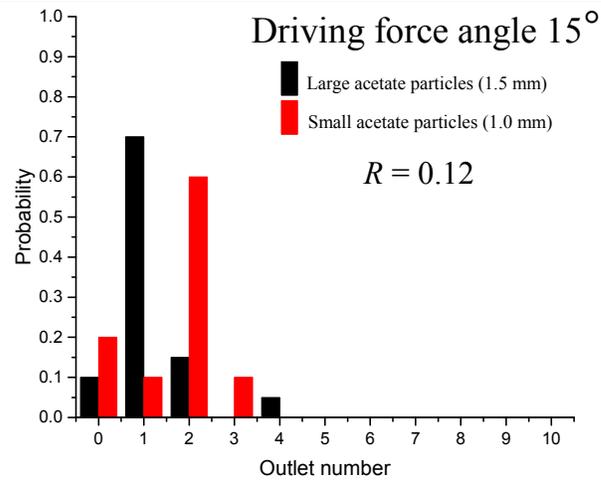
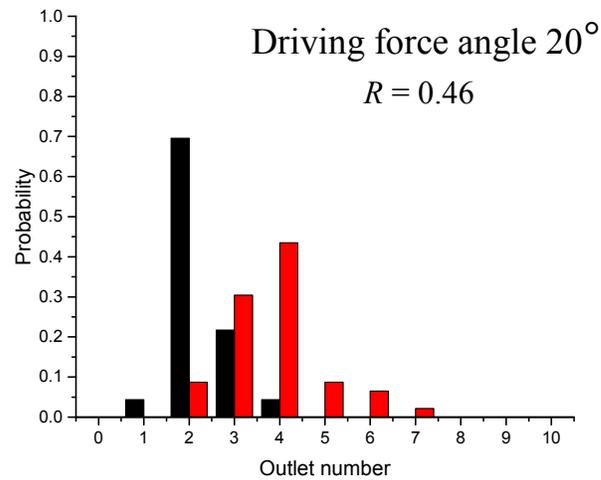
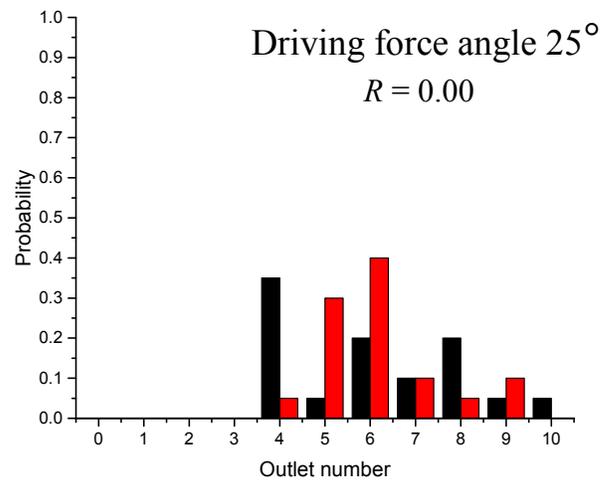

**Figure 5:** Probability distributions of particles at the outlet for different forcing angles, as indicated. The sample is a binary mixture of cellulose acetate particles (1.0 mm and 1.5 mm in diameter).

Then, in order to reduce, in principle, the spread in the distribution of particles we performed experiments in which only two particles are released simultaneously in each experiment. In addition, to further demonstrate the negative effect that an increase in the initial number of particles has on the resolution of the separation, we also performed experiments increasing the initial number of particles in the inlet of the device.

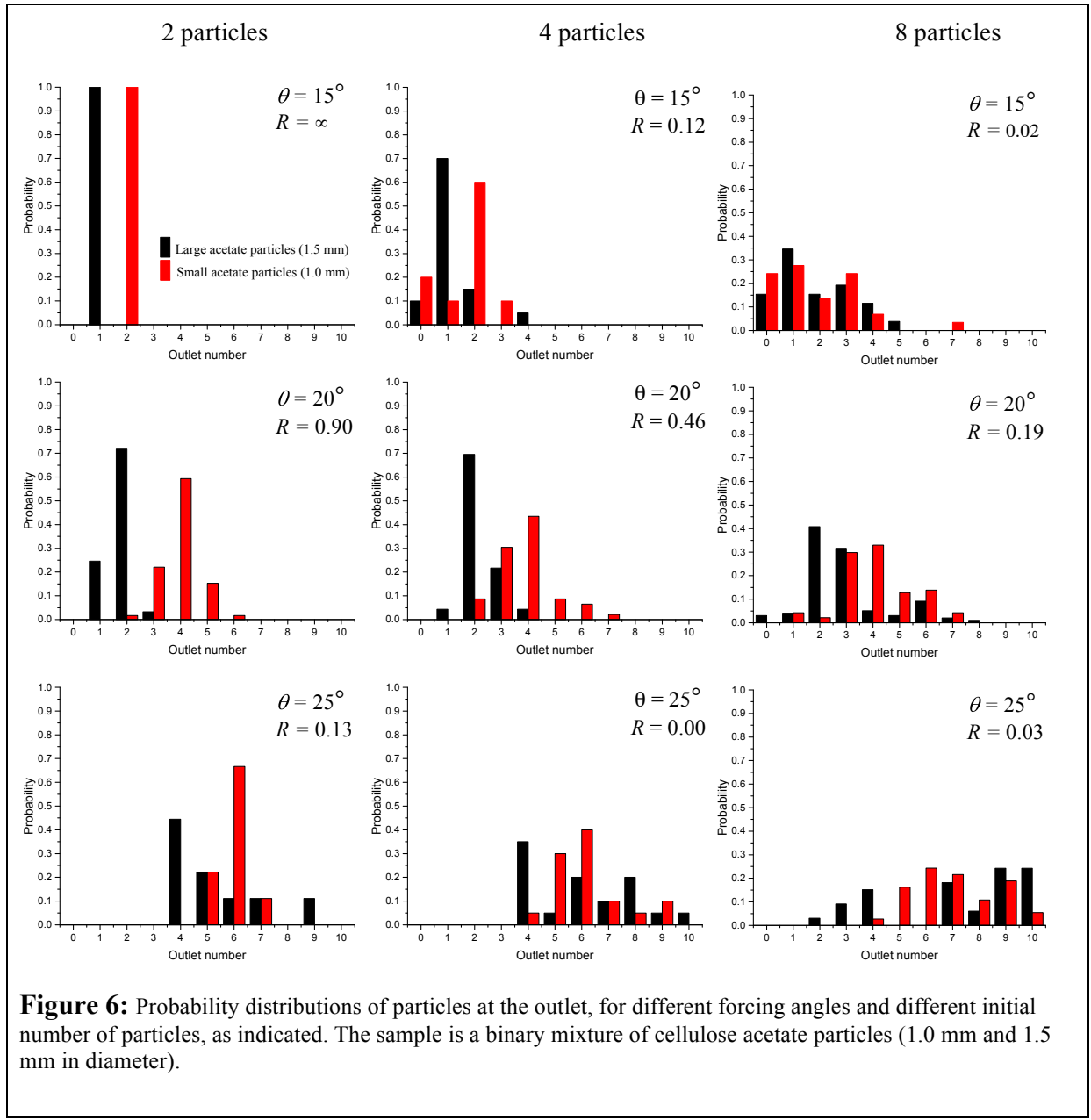

**Figure 6:** Probability distributions of particles at the outlet, for different forcing angles and different initial number of particles, as indicated. The sample is a binary mixture of cellulose acetate particles (1.0 mm and 1.5 mm in diameter).

In Figure 6, we present the distributions of cellulose acetate particles at the outlet corresponding to experiments in which particles are released in groups of 2, 4 and 8, as indicated. We also investigated different forcing angles as before, $\theta = 15°$, 20° and 25°. First of all, it is clear that for any forcing angle, the resolution decreases with the number of particles released simultaneously (plots from left to right in the figure). This is probably due to a combination of particle-particle interactions as well as the fact that larger groups of particles result in larger random variations in the initial position of the particles at the inlet. It is also clear from the figure that, in general, the optimal resolution is obtained at the intermediate forcing angle. At these intermediate angles, big particles are restricted to the lower output bins due to locking, whereas the small particles migrate at larger angles and reach the central outlets. It is also important to point out that the case of two particles driven at $\theta = 15°$ is special, in that the behavior seems completely deterministic, with zero dispersion and infinite resolution. However, the results obtained with larger groups of particles in the inlet area suggest that this might not be an ideal condition, in that any increase in the initial number of particles could dramatically reduce the resolution. This is probably due to the fact that the initial separation was only between adjacent bins (outlets one and two), a situation in which even a small perturbation in the trajectory of the particles could have a big effect on the ability to separate them.

## Conclusion

We performed experiments in a scaled-up model of DLD microfluidic devices in which suspended particles move through a square array of cylindrical posts under the action of centrifugation. Based on previous studies we designed the posts to be smaller than the size of the particles to be separated. Experiments demonstrated the fractionation of binary suspensions of spherical particles depending on the forcing angle. We measured the resolution of separation by collecting the suspended particles in a discrete set of bins located at the outlet. We found excellent resolution when large particles are locked to move at small migration angles, without moving across columns of obstacles, but smaller particles follow the direction of centrifugal acceleration more closely. Finally, we showed that resolution improves as the initial concentration of particles decreases.

The observed transport and fractionation of suspended particles under centrifugal force is analogous to the behavior exhibited in other force-driven DLD systems, including gravity- and electrokinetically-driven cases. Although further work at the micro scale is needed, the results presented here indicate that models used to describe the motion in those systems can be extended to centrifuge-driven DLD (c-DLD). In addition, previous agreement between micro and meso scale systems suggests that our experiments could inform future development in c-DLD systems.

## Acknowledgements

This work was partially supported by the National Science Foundation Grant No. CBET-1339087, along with support from the Rutgers School of Engineering, the Department of Mecahnical and Aerospace Engineering, and an A. Walter Tyson Assistant Professorship Award. We also thank Maxim Lazoutchenkov, Brett Cecere, Maxwell Berka, Derek Dechent, Kshitij Minhas, Mark Mettias, and Joseph Stephens for working on the design of the device as part of their senior design course. Finally, we acknowledge Saugata Dutt, Sandesh Gopinath, and Chen Yang for their assistance with the experimental setup